\documentclass[%
 reprint,
 amsmath,amssymb,
 aps,
 prab,
]{revtex4-2}

\usepackage{glossaries}
\usepackage{hyperref}
\usepackage[capitalise]{cleveref}
\usepackage{graphicx}
\usepackage{bm}
\usepackage{xcolor,soul}
\usepackage{siunitx}

\newacronym{pic}{PIC}{particle-in-cell}
\newacronym{pwfa}{PWFA}{plasma wakefield accelerator}
\newacronym{smi}{SMI}{self-modulation instability}
\newacronym{dps}{DPS}{discharge plasma source}

\begin{document}

\title{Influence of ion motion in a resonantly driven wakefield accelerator}

\author{Erwin Walter}
\affiliation{Max Planck Institute for Plasma Physics, 85748 Garching, Germany}
\affiliation{Exzellenzcluster ORIGINS, 85748 Garching, Germany}
\author{John P. Farmer}
\email{j.farmer@cern.ch}
\affiliation{Max Planck Institute for Physics, 80805 Munich, Germany}%
\author{Marlene Turner}%
\affiliation{CERN, 1211 Geneva 23, Switzerland}
\author{Frank Jenko}
\affiliation{Max Planck Institute for Plasma Physics, 85748 Garching, Germany}

\begin{abstract}
Several different schemes for plasma wakefield acceleration using a train of drivers have been pursued, based on the resonant excitation of a plasma wave. Since these schemes rely on the plasma electron wave surviving for many periods, the motion of the plasma ions can have a significant impact on the beam--plasma interaction. In this work, simulations are used to study the impact of this ion motion on the development of the self-modulation of a long beam, directly applicable to recent experiments. It is shown that two related but distinct effects contribute to the suppression of the wakefield excitation: the loss of resonance between the drive beam and the plasma wave it excites, and phase mixing due to transverse wavebreaking. Although only the latter has previously been investigated, we show that the two effects follow the same scaling with ion mass.
\end{abstract}

\maketitle

\section{Introduction}
Plasma wakefield acceleration continues to attract significant interest because of the large fields which plasma can support. A driver, either a laser pulse~\citep{Tajima1979} or a charged particle bunch~\citep{Chen1985}, excites a plasma wave, which in turn is used to accelerate a trailing witness bunch. Several schemes have been proposed and demonstrated based on resonant excitation, in which a periodic train of drivers acts to resonantly excite the plasma wave, with each driver in the train contributing to the total wakefield amplitude. In this case, the drive beam may either be premodulated~\citep{Clayton1993,Jakobsson2021}, or rely on the exploitation of beam--plasma instabilities to ``self-modulate'' a long beam, causing it to break up into a train of short drivers with a frequency equal to the plasma frequency~\citep{Andreev1992,Krall1993,Lotov1998}.

Maintaining the resonance between the drive beam and the plasma wave it excites is a key consideration for periodic drivers. At high wakefield amplitudes, the nonlinear shift in the plasma frequency modifies the resonance condition, and the resulting dephasing can result in the saturation of the wakefield amplitude~\citep{Rosenbluth1972,Lotov2013}. The radial variation in wakefield amplitude also results in a bowing (curvature) of the wakefields~\citep{Esarey2009}.  For the case of a premodulated driver, this shift in resonance can be compensated by varying the periodicity along the length of the drive beam~\citep{Umstadter1994,Chiadroni2017,Wetering2024}. For a self-modulated driver, wakefields grow both along the length of the driver and along its propagation length~\citep{Kumar2010,Schroeder2011}, leading to a resonance condition with a spatio-temporal dependence. The effect can, however, be partially compensated by a plasma density step after the initial phase of self-modulation~\citep{Lotov2015}, or avoided by injecting multiple witness bunches to limit the wakefield amplitude~\citep{Farmer2024b}.

In addition to these effects, the ponderomotive-like force due to the oscillating electric fields will act on the plasma ions~\citep{Silva1999,Gorbunov2003,Gilljohann2019}:
\begin{equation}
F_\mathrm{p,i}=-\frac{Ze^2}{4m\omega_p^2}\nabla\hat{\bm{E}}^2 
\end{equation}
with $Ze$ the ion charge, $m$ the electron mass, $\omega_p$ the plasma frequency, and $\tilde{E}$ the wakefield envelope.  Since the wakefields grow along the length of the drive beam, reaching the large gradients necessary for acceleration requires that the resonance between the driver and plasma be maintained over many plasma periods. These relatively long timescales mean that significant ion motion may develop for sufficiently light plasma ions. 

The plasma wave grows over many plasma periods and has a phase velocity approximately equal to the velocity of the drive beam, resulting in a time-averaged force on the plasma ions which is predominantly transverse.  The resulting radial variation of the ion density results in a corresponding shift in the local plasma frequency. Such transversely inhomogeneous plasma waves are known to lead to transverse wavebreaking~\citep{Bulanov1997}, which occurs as the laminar oscillations of the plasma electrons cross~\citep{Dawson1959}. The resulting phase mixing leads to a strong damping of the plasma wave~\citep{Gorbunov2003,Spitsyn2018}, which in turn suppresses the self-modulation of the proton beam tail~\citep{Vieira2012,Vieira2014}.

In this work, we investigate the role of ion motion in the self-modulation instability, with parameters chosen to reflect recent experiments carried out at CERN by the AWAKE project~\citep{Turner2024}.
It is shown that ion motion leads to two related but distinct effects which influence the beam--plasma interaction: the damping of the plasma wave due to transverse \emph{wavebreaking}, as discussed in previous works; and the enhanced \emph{detuning} between the drive beam and the plasma wave it excites, which leads to a foreshortening of the microbunch train. It is shown that for the parameters of interest, detuning dominates in the early stage of the instability, near the beam head, while wakebreaking dominates towards the beam tail. 
We show that the newly identified effect of detuning due to ion motion follows the same scaling with ion mass as found by \citet{Spitsyn2018} for wavebreaking.



\section{Applicability to experiment}
The AWAKE experiment at CERN makes use of proton beams to drive plasma wakefields. The high energy of protons which can be achieved in existing synchrotrons, 400~\si{\giga\electronvolt} in the case of the SPS beam used in AWAKE, have the potential to drive wakefields over much longer distances than electron or laser drivers~\citep{Lotov2021}, allowing a witness bunch to reach high energy in a single acceleration stage.

Although schemes to provide short proton beams are currently being investigated~\citep{Farmer2024}, currently available beams are too long to effectively drive a wake, spanning hundreds of plasma periods. The wakefields driven by different sections of the drive beam destructively interfere, resulting in low-amplitude waves not suitable for particle acceleration. However, these fields are sufficient to provide periodic focusing and defocusing of the drive beam~\citep{AWAKE2019}. AWAKE harnesses this ``self-modulation'' to generate a train of microbunches which can resonantly excite large amplitude wakefields suitable for acceleration~\citep{AWAKE2018}.

In order to exploit the high energy of the proton driver, the AWAKE project is pursuing technologies which have the potential to generate long (of the order of \si{100~\metre}) plasma sections~\citep{
AWAKE2022}. As part of this effort, an experimental run using a \gls{dps} was carried out~\citep{torrado2023}. This novel technology allowed several studies which were not previously possible in AWAKE, such as the filamentation of an initially wide drive beam~\citep{Verra2024,Walter2024}. The \gls{dps} also supported operation with different noble gases, with experimental measurements of the proton beam profile after the plasma carried with xenon, argon and helium~\citep{Turner2024}. By varying the gas used in the discharge, the influence of ion motion on the self-modulation instability could be studied.

Simulations were carried out in support of the experimental campaign, providing additional insights into the plasma dynamics which were not directly observed in the experiment. The simulations were then validated through comparison to the time-resolved images from the experiment. The axisymmetric particle-in-cell code LCODE~\cite{lotov2003,lcode-manual} was used to carry out the simulation studies. The parameters were chosen to reflect the experiment, with a \si{49\,\nano\coulomb}, \si{400\,\giga\electronvolt} proton drive beam, modelled as being initially Gaussian with an RMS length of \si{170\,\pico\second}, a transverse size of \si{160\,\micro\metre}, and a normalized emittance of \si{\num{2.2}\,\micro\metre}, propagating through \si{10\,\metre} of plasma with a number density of \si{\num{4.8e14}\,\centi\metre^{-3}}.

\begin{figure}
 \includegraphics[width=\linewidth]{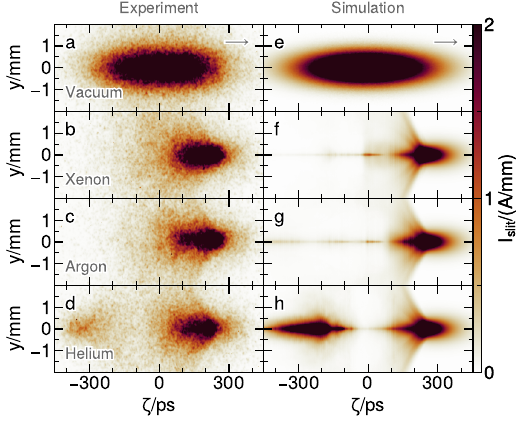}
 \caption{\label{fig:ion_expstreak}Time-resolved proton beam profiles from experiment (a-d) and simulation (e-h) after propagating for $10\,\mathrm{m}$ in vacuum (a,e), or a plasma of ionized xenon (b,f), argon (c,g) or helium (d,h). The images are captured using a streak camera positioned 3.5\si{\,\metre} after the plasma source. For a full discussion of the experimental results, please refer to \citet{Turner2024}.}
\end{figure}

The experimental and simulation results are shown in \cref{fig:ion_expstreak}. In all cases shown here, the plasma density and the initial beam parameters are kept constant, and only the choice of plasma species is changed: no plasma (vacuum), xenon, argon and helium. A detailed analysis of the experimental results is carried out in \citet{Turner2024}, including additional studies varying both the beam population and the plasma density. In the present work, we limit the analysis of the experimental results to identifying the key signatures of the self-modulation instability and the influence of ion motion upon it. Demonstrating that simulations reproduce these effects motivates the simulation studies which constitute the bulk of this paper.

In the experiment, time-resolved images of the proton beam were captured using a streak camera~\citep{Campillo1983} placed \si{\num{3.5}\,\metre} after the plasma exit. The simulated proton beam is propagated by an equivalent distance, transformed into Cartesian coordinates, and cropped according to the streak-camera aperture, an \si{80~\micro\metre}-wide slit. When used to image the entire length of the beam, the temporal resolution of the streak camera is insufficient to resolve the individual microbunches, which have a periodicity of $\sim5$\si{\,\pico\second}, and so only the charge envelope is measured. An equivalent smoothing is applied to the simulation results to facilitate a comparison. Both experimental and simulation images are saturated to emphasize the structure of the beam.

The proton beam is focused at the entrance of the \gls{dps}, and, in the absence of plasma, reaches an RMS width of \si{460~\micro\metre} at the streak camera \si{\num{13.5}~\metre} downstream, shown in \cref{fig:ion_expstreak}a,e.  When plasma is present, the proton beam is self-modulated.

In the early stage of the instability, the wakefields are essentially purely focusing, as the initially-at-rest plasma electrons are pulled towards the beam axis, and oscillate under the restoring force of the plasma ions. This periodic focusing of the beam reinforces the wakefields, and the instability grows. As the wakefield amplitude increases, periodically defocusing regions develop. Since the instability grows both along the beam and along the length of the plasma,  this net focusing is therefore limited to the beam head (the leading portion of the beam, $\zeta>0$, towards the right of the plots), and characterised by a decrease of the observed beam radius compared to the case with no plasma.  The development of periodic defocusing regions can be observed by the rapid broadening of charge envelope at $\zeta\approx200$~\si{\pico\second}.

The periodic focusing and defocusing of the drive beam impacts the streak camera images in two ways. Defocusing leads to a reduction of the beam charge density while still inside the plasma. Perhaps counterintuitively, strong focusing also reduces the detected charge, as tightly-focused microbunches diverge more rapidly in the region between the plasma and the streak camera~\citep{Moreira2018}.  For the case of xenon, shown in \cref{fig:ion_expstreak}b,f, these effects result in only the head of the proton beam reaching the streak camera. Decreasing the mass of the plasma ions increases the degree of ion motion, allowing its influence on the beam--plasma interaction to be understood. For argon (\cref{fig:ion_expstreak}c,g), no significant change to the beam profile is apparent, although the total charge reaching the streak camera is slightly increased. For helium (\cref{fig:ion_expstreak}d,h) the formation of a tail (the trailing part of the beam) is clearly visible, arising as ion motion acts to suppress the self-modulation instability, with the result that more charge reaches the streak camera.

The simulation results reproduce all of the behaviour observed in the experimental data, providing confidence that the relevant physical processes are being correctly modelled. We note from  that the formation of the tail is stronger in simulation than experiment. This occurs because the self-modulation in the DPS run was \emph{unseeded}~\cite{Batsch2021,Verra2022} with the instability growing from fluctuations in the beam profile. Computational constraints on the number of simulated macroparticles result in a beam which is comparatively more granular than in experiment, and so these fluctuations are larger. This results in simulated wakefields which grow from a larger initial amplitude, reaching high amplitude earlier along the beam.  The ion motion driven by these wakefields therefore also develops sooner, resulting in the longer beam tail. While this difference between simulation and experiment leads to differences in the absolute time for effects to develop, the scaling of the impact of physical quantities on ion motion and related effects should be preserved.

\section{Plasma dynamics}\label{sec:instab}
\begin{figure*}[ht]
 \includegraphics{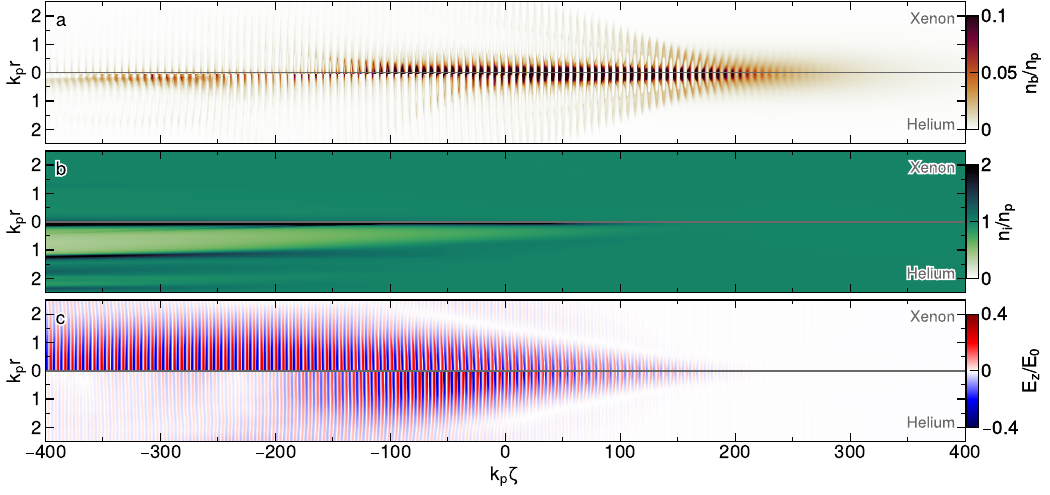}
 \caption{\label{fig:ion_fields}The plasma response to a proton beam undergoing self-modulation after a propagation distance of $s=$\si{5\,\metre}. a) The beam density, b) the plasma ion density, and c) the longitudinal component of the wakefield are shown for xenon (top) and helium (bottom) plasmas.}
\end{figure*}


As discussed above, simulations allow the beam--plasma interaction to be followed throughout its evolution, and can investigate quantities which are not easily accessible in experiment.  The plasma electrons respond much faster than the relativistic proton driver, and so over short distances the driver can be considered ``frozen'' as it passes through the plasma. \Cref{fig:ion_fields} shows a snapshot of the proton beam and the associated plasma response after \si{5\,\metre} of plasma propagation ($k_pz=2\times10^5$, with $1/k_p=c/\omega_p$ the plasma skin depth), halfway through the \gls{dps}.  

From top to bottom, \cref{fig:ion_fields} shows a) the beam density, b) the plasma ion density, and c) the longitudinal wakefields.  The top half of each plot shows xenon, the heaviest ion used in the experiment, while the bottom half shows helium, the lightest.  For helium, the plasma ion density, shown in \cref{fig:ion_fields}b, is significantly perturbed.  The symmetry of the system requires that the envelope of transverse electric field is zero on axis leading to a time-averaged inward force for ions close to the axis, while ions further from the axis are pushed outwards, leading to a peak in the ion density on axis, surrounded by an annular region of lower ion density.  This variation in the ion density perturbs the plasma wakefields, shown in \cref{fig:ion_fields}c, which in turn imprint back on the proton beam, shown in \cref{fig:ion_fields}a.  For xenon, the plasma ion density remains essentially uniform, and so comparing the beam--plasma interaction for the two ion species allows the impact of ion motion to be investigated.

\begin{figure}
 \includegraphics[width=\linewidth]{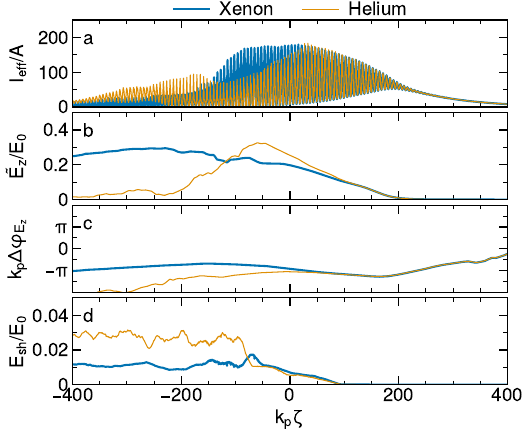}
 \caption{\label{fig:ion_fields2}Lineouts showing key metrics of the beam--plasma interaction after \si{5\,\metre} propagation in xenon and helium plasmas.  a) The effective current of the proton driver, $I_\mathrm{eff}$; b) the envelope, $\tilde{E_z}$, and c) the relative phase, $\Delta\varphi\left(E_z\right)$, of the on-axis longitudinal wakefield; and d) the plasma sheath field, evaluated at $k_pr=10.25$.
 }
\end{figure}

Rather than comparing the full two-dimensional plots in \cref{fig:ion_fields}, it is convenient to instead consider quantities that depend only on $\zeta$.  The proton microbunch train can be characterised by its effective current,
\begin{equation}
  I_\textrm{eff}\left(\zeta\right)=\frac{c}{\Delta\zeta}\int_\zeta^{\zeta+\Delta\zeta} q \operatorname{K}_0\left(k_pr\right)
\end{equation}
with $c$ the vacuum speed of light, $\operatorname{K_0}$ the zeroth-order modified Bessel function of the second kind, $\Delta \zeta$ some arbitrary interval chosen to be much shorter than $1/k_p$,
and $q$ and $r$ the charge and radial position of all particles (macroparticles) in this range.  The effective current is a measure of the longitudinal wakefield amplitude excited by each slice $\Delta\zeta$ of the drive beam~\citep{Katsouleas1987}.  In the absence of self-modulation, the wakefields excited by different slices destructively interfere, leading to a low wakefield amplitude.  The effective current of the proton beam after 5~\si{\metre} of plasma propagation is shown in \cref{fig:ion_fields2}a for both the helium and xenon cases.

The longitudinal wakefield, $E_z$, can be characterised in terms of its on-axis amplitude and phase, shown in \cref{fig:ion_fields2}b,c.  The amplitude is represented as the envelope $\tilde{E_z}$, chosen as the magnitude of the trough minima, corresponding to the accelerating gradient for an electron witness bunch, and represented in terms of the cold wavebreaking amplitude, $E_0=mc^2k_p/e$, with $m$ the electron mass.  The phase is characterized by the phase difference relative to a harmonic wave with wavelength $2\pi/k_p$, evaluated at the zero crossings, i.e.
\begin{equation}
    \Delta\varphi_{E_z}\left(\zeta_{\mathrm{zero},n}\right)=\zeta_{\mathrm{zero},n}+n\frac{\pi}{k_p} - \zeta_{\mathrm{zero},1}
\end{equation}
with $\zeta_{\mathrm{zero},n}$ the position of the $n$th zero-crossing.

Finally, the onset of wavebreaking, which has previously been identified as a key mechanism through which ion motion impacts on the plasma wakefields, can be identified by the development of a radial sheath field, $E_\mathrm{sh}$ around the plasma~\citep{Gorn2020}, shown in \cref{fig:ion_fields2}d.  This field develops as the trajectory crossing associated with wavebreaking allows the exchange of energy between plasma electrons, causing some to be ejected transversely from the plasma.

Taken together, the simplified plots in \cref{fig:ion_fields2} show the key aspects of the plasma response, plus some additional features which are not readily apparent from \cref{fig:ion_fields}.  The same initial beam distribution was used for both simulations, ensuring that, in the absence of ion motion, self-modulation will develop identically in both cases.  Since ion motion is driven by the ponderomotive action of the wakefields, which themselves grow along the length of the drive beam, the plasma response is identical near the beam head ($k_p\zeta\geq100$).  Beyond this point, the perturbation to the ion density is sufficient to modify the plasma reponse.  

For the case of helium, ion motion first results in an increase in the longitudinal wakefield amplitude, starting around $k_p\zeta=100$.  This effect has been discussed by \citet{Minakov2024}, and is attributed to energy transport within the plasma.  This increase in wakefield amplitude is only transient, with the wakefields strongly damped towards the beam tail.   There is also an evident shift in the wakefield phase for the case of helium, as the reduction in the plasma ion density leads to a decrease in the local plasma frequency.  This phase shift leads to the erosion of the drive beam, as microbunches fall out of phase with the wakefields and are defocused, reducing the effective current.

The sheath field for both xenon and helium are initially very similar, appearing at the same position along the beam, and with a similar magnitude.  However, in the trailing half of the drive beam ($\zeta<0$), the sheath field is significantly enhanced for helium.  This indicates that wavebreaking occurs at the same position in both cases,  but that the additional bowing of the wakefields due to ion motion in helium results in significantly more energy being extracted from the plasma wave due to phase mixing, as borne out by the strong damping of the wakefields towards the beam tail ($k_p\zeta\leq-100$).  Finally, this suppression of the wakefields also reduces the focusing and defocusing fields (not shown) acting on the proton beam, suppressing self-modulation, with the result that the tail of the proton beam survives in the case of helium.  For xenon, there is no significant ion motion, and the wakefields extend the entire length of the beam.  This leads to the loss of the proton beam tail due to the phase shift associated with plasma nonlinearity.


\section{Instability growth}
While the simulation snapshot shown in \cref{fig:ion_fields,fig:ion_fields2} allow the plasma dynamics to be understood, the proton beam retains the imprint of the wakefields at earlier propagation distances, and so depends on the integrated plasma response over the plasma propagation up to that point.  This spatiotemporal dependence of SMI means that both the growth along the beam and the growth along the plasma should be considered in order to gain a complete understanding of the beam--plasma interaction.

\begin{figure}
 \includegraphics[width=\linewidth]{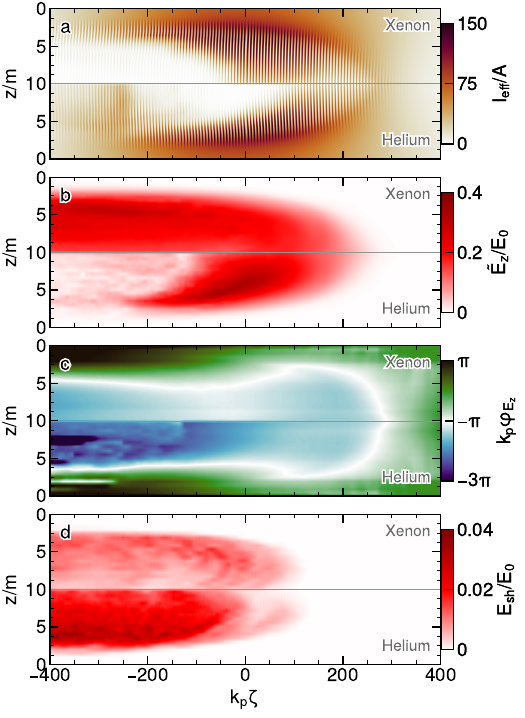}
 \caption{\label{fig:ion_watfal}Spatiotemporal evolution of the proton beam and plasma response over \si{10\,\metre} of plasma interaction, for xenon (upper of each plot) and helium (lower of each plot).  Right to left shows the evolution along the beam, with the beam propagating to the right, while upper/lower edge towards the centre shows the evolution along the propagation length.  a) the beam effective current; b) the amplitude and c) the relative phase of the longitudinal wakefield; and d) the plasma sheath field.}
\end{figure}


Since the quantities in \cref{fig:ion_fields2} are functions only of $\zeta$, their evolution along the length of the plasma can be displayed as ``waterfall'' plots, shown in \cref{fig:ion_watfal}.  As in \cref{fig:ion_fields}, the top half of each plot corresponds to xenon, the bottom half to helium.  The proton beam and the associated plasma response at the plasma entrance are shown at top/bottom of the respective plots, with the progression towards the centre of the plot showing the evolution along the 10~\si{\metre} plasma length.  The centre of the plots correspond to the beam and plasma response at the plasma exit.

The plots show the spatiotemporal nature of several key aspects of the SMI instability, and highlight the influence of ion motion.  The initial proton beam in the two cases is identical, and the excited plasma wave is too small to drive significant ion motion, and so the initial plasma response and the proton bunch evolution are the same in both cases.  The plasma response leads to periodic focusing of the proton beam, increasing the effective current, and increasing the amplitude of the excited plasma wave.  As the wakefields grow in amplitude, defocusing regions develop, leading to the development of the microbunch train.  Since the instability grows in both space and time, the microbunch formation starts near the back of the proton beam and advances towards the beam head as it propagates through the plasma.

Unlike the case after 5~\si{\metre} plasma propagation, shown in \cref{fig:ion_fields2}, where wavebreaking occurred for both ion species at the same point, it can be seen that at short propagation distances, $z\approx1~\si{\metre}$, wavebeaking occurs for helium but not for xenon.  At later propagation distances, wavebreaking occurs at the same position along the beam, but the magnitude of the sheath field is greatly enhanced for the case of helium, accompanied by strong damping of the wakefields towards the beam tail.  As the instability grows, the region of strong wakefield damping moves forwards in the beam frame, but remains constrained to the trailing half of the beam.

For both helium and xenon, the phase of the wakefields evolve as the instability grows, leading to variations both along the beam and along the plasma, as seen in \cref{fig:ion_watfal}c.  The wakefields early in the instability determine the phase of the microbunch train, and as the wakefield phase evolves, microbunches fall into the defocusing phase of the wakefield and are lost.  The additional phase shift due to ion motion causes this process to happen sooner, leading to a shorter microbunch train in the case of helium.  However, since ion motion also results in enhanced wavebreaking, the tail of the beam survives for helium.



\section{Scaling with ion mass}
A simple analytic argument for the timescale for wavebreaking to occur was laid out by~\citet{Spitsyn2018}.  The wakefields exert a force on the plasma ions, and so the time (or, in the beam frame, the distance in $\zeta$) for an ion to move a given distance scales as the root of the acceleration, i.e.\ $\sim \smash{m_i^{-1/2}}$, which determines the timescale for the variation in the ion density, as also shown by~\citep{Vieira2014}.  For small density perturbations, the shift in the local plasma frequency can be approximated by the first-order Taylor expansion, and so has the same scaling.  The time for a given phase advance depends on the integral of the perturbed plasma frequency, and therefore scales as $\smash{m_i^{-1/3}}$.  Wavebreaking occurs when the trajectories of the plasma electrons cross, and will occur when the phase difference between plasma electron oscillations at different radii approaches $\pi$.  Wavebreaking, and the associated strong damping of the plasma wave, therefore follows this $\smash{m_i^{-1/3}}$ scaling.

Detuning, which dominates near the head of the drive beam, depends on the absolute phase shift of the plasma oscillations, as the proton microbunches fall out of resonance with the wakefields they excite.  Ultimately, they fall into the defocusing phase of the wakefields and are lost.  Since this effect also arises due to a phase shift, it should display the same $\smash{m_i^{-1/3}}$ dependence.

\begin{figure}
 \includegraphics[width=\linewidth]{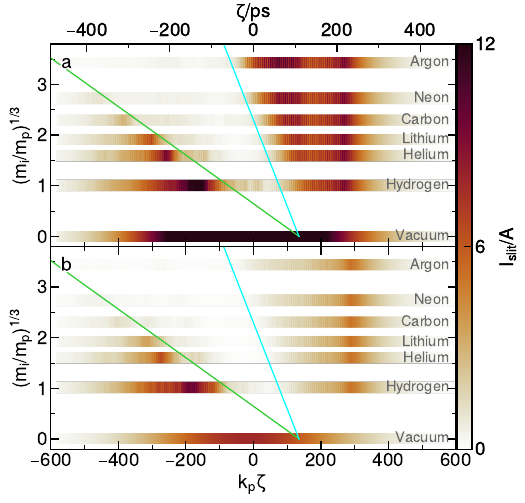}
 \caption{\label{fig:ion_ymscal}Current profiles of the proton beam passing through a virtual aperture for different plasma species a) at the plasma exit ($z=10\,\mathrm{m}$) and b) after propagation to a screen placed 3.5~\si{\metre} after the plasma exit, as in \cref{fig:ion_expstreak}e-h. The cyan line indicates the end of the beam head, and the green line indicates the start of the tail.}
\end{figure}

To confirm this scaling, simulations were carried out with a range of ion masses.  The proton beam is propagated through 10~\si{\metre} of plasma, and then an additional 3.5~\si{\metre} through vacuum, as in the experiment.  In each case, the current passing through a virtual aperture, as used for the streak camera comparison in \cref{fig:ion_expstreak}, is calculated, with the same smoothing applied.  The results, both at the plasma exit and after vacuum propagation, are shown in \cref{fig:ion_ymscal}.  Propagation with no plasma, i.e.\ 10 (a) or 13.5~\si{\metre} (b) of vacuum propagation, is shown for comparison.

Bands representing the current profile are shown for each plasma species, with bands spaced in the vertical direction according to $\smash{m_i^{1/3}}$.  A timescale that varies as $\smash{m_i^{-1/3}}$ should therefore appear as a straight line in the plot.  
Guide lines have been added for the foreshortening of the proton beam head (cyan) and the appearance of the beam tail (green).  In the case of vacuum propagation, there is no ``gap'' in the proton beam current profile due to self-modulation, and so the additional constraint was added that the two guide lines should meet at $\smash{m_i}=0$.  As can be seen from the plot, the impact of both detuning and wavebreaking follow the expected $\smash{m_i^{-1/3}}$ dependence, both at the plasma exit and after vacuum propagation.  The foreshortening of the microbunch train is less visible after vacuum propagation due to the stronger divergence of the trailing microbunches, which could explain why it was not readily apparent from the experimental images shown in \cref{fig:ion_expstreak}.

\section{Conclusion}
The self-modulation of a long beam in plasma in the presence of ion motion has been investigated through the use of particle-in-cell simulations, with parameters chosen to be relevant to experiments carried out at AWAKE~\citep{Turner2024}.  It is shown that ion motion impacts the growth of SMI through two distinct mechanisms, both linked to the shift in resonant frequency of the plasma wave.  In the absence of ion motion, nonlinearities in the plasma response lead to the microbunch train falling out of resonance with the wakefields it excites, ultimately leading to the loss of trailing microbunches as they fall into the defocusing phase of the wakefields.  The additional frequency shift due to ion motion results in this dephasing occurring sooner, leading to a shorter microbunch train.  Further back in the beam, the transverse variation of the local plasma frequency due to ion motion leads to enhanced wavebreaking, strongly damping the plasma wave.  This damping allows the tail of the drive beam to survive, even as the microbunch train near the beam head is foreshortened.

Previous studies have only investigated the influence of ion motion on SMI through the wavebreaking mechanism.  In this work, it is shown that since both effects arise due to a phase shift of the plasma wave, the newly discovered effect of detuning follows the same $\smash{m_i^{-1/3}}$ scaling as previously derived for wavebreaking.

Understanding the mechanisms through which ion motion impacts on the self-modulation of a drive beam is vital for the planning of future experiments based on this technique, and will allow additional constraints to be placed on the choice of ion for future AWAKE experiments. 



\begin{acknowledgments}
We are grateful to the referees of the experimental manuscript by \citet{Turner2024} who highlighted the scaling with ion mass as a key topic.
\end{acknowledgments}

\section*{Author contribution statements}
This work was concieved by MT in support of the experimental study published in \citet{Turner2024}. The investigation and visualisation were carried out by EW, with supervision from JPF.  Methodology and manuscript by JPF.  FJ contributed to funding acquisition and provided additional supervision.

\bibliography{ion}{}

\end{document}